\documentclass{elsart}
\usepackage{graphicx,epsfig,natbib,amssymb}

\newcommand{\gsim}{\mbox{\hspace{.2em}\raisebox{.5ex}{$>$}\hspace{-.8em}\raisebox{-.5ex}{$\sim$}\hspace{.2em}}}

\newcommand{\ssst}{\scriptscriptstyle}
\newcommand{\E}[1]{\times 10^{#1}}

\newcommand{\s}{\,{\rm s}}      \newcommand{\ps}{\,{\rm s}^{-1}}
\newcommand{\yr}{\,{\rm yr}}    
\newcommand{\cm}{\,{\rm cm}}    
\newcommand{\parsec}{\,{\rm pc}}\newcommand{\kpc}{\,{\rm kpc}}
\newcommand{\ergs}{\,{\rm ergs}}        \newcommand{\K}{\,{\rm K}}
    \newcommand{\keV}{\,{\rm keV}}

\newcommand{\nel}{n_{e}}        \newcommand{\NH}{N_{\ssst\rm H}}
\newcommand{\no}{n_{\ssst 0}}

\newcommand{\rs}{r_{s}}         \newcommand{\vs}{v_{s}}
\newcommand{\nH}{n_{\ssst\rm H}}

 \newcommand{\ASCA}{{\sl ASCA}}
\newcommand{\Chandra}{{\sl Chandra}}
\newcommand{\du}{d_{8}}         \newcommand{\Eu}{E_{51}}

\begin{document}

\begin{frontmatter}

\title{Highly Clumpy Structure of the Thermal Composite Supernova Remnant 3C~391 Unveiled by \Chandra \thanksref{fund}}
\thanks[fund]{
Supported by NSFC grants 10073003 \& 10221001,
CMST grant NKBRSF-G19990754,
NASA contract NAS8-39073, and NASA grants GO2-3081X \& NAG5-8935.
}

\author{Yang Chen\thanksref{NJU}}, \author{Yang Su\thanksref{NJU}},
 \author{Patrick O.\ Slane\thanksref{CfA}},
 \author{Q.\ Daniel Wang\thanksref{UMass}}
\thanks[NJU]{Department of Astronomy, Nanjing University, Nanjing 210093,
       P.R.China}
\thanks[UMass]{Harvard-Smithsonian Center for Astrophys.,
Cambridge, MA 02138}
\thanks[CfA]{Dept.\ of Astr., B619E-LGRT, 
       Univ.\ of Massachusetts, Amherst, MA01003}

\begin{abstract}
The nature of the internal thermal X-ray emission
seen in ``thermal composite" supernova remnants is still uncertain.
\Chandra\ observation of the 3C391 shows a southeast-northwest
elongated morphology
and unveils a highly clumpy structure of the remnant.
Detailed spatially resolved spectral analysis for the small-scale
features reveals normal metal abundance and uniform temperature
for the interior gas.
The properties of the hot gas comparatively favor the cloudlet evaporation
model as a main mechanism for the ``thermal composite" X-ray appearance,
though radiative rim and thermal conduction may also be effective.
A faint protrusion is found in Si and S lines out of the southwest
radio border.

\end{abstract}

\end{frontmatter}

\section{Introduction}
3C~391 (G31.9+0.0), with irregular morphology (e.g.\ Chen \& Slane 2001), 
has been classified into the ``thermal composite''
category of supernova remnants (SNRs)
(Rho \& Petre 1998).
They generate bright thermal X-ray emission interior to
their radio shells, and have faint X-ray rims.
They are usually found to interact with
adjacent molecular clouds, characterized by the hydroxyl radical
maser emission (e.g.\ Yusef-Zadeh et al.\ 2003).
The nature of the internal thermal X-ray emission
seen in mixed morphology remnants is still uncertain.

So far at least four candidate scenarios compete to account for
centrally-brightened X-ray morphology.
The first scenario is radiative cooling of the rim gas.
Under this hypothesis, the gas at the rim has been cooled down in the
radiative stage, with a temperature so low that its
X-ray emission is very weak, while the gas in the inner volume
is still hot enough to emit strong X-rays (e.g.\ Rho \& Petre 1998).
The second mechanism invokes thermal conduction.
It is suggested that thermal conduction
in the remnant can prevent formation of the very tenuous, hot gas
in the inner part
and therefore change the interior structure from the standard
Sedov solution, resulting in a nonnegligible density and luminous
X-ray in the interior (as observed in, especially, the radiative stage)
(Cox et al.\ 1999).
The third scenario is cloudlet evaporation in the SNR interior.
When an SNR expands in an inhomogeneous interstellar medium (ISM)
whose mass is mostly contained in small clouds,
the clouds engulfed by the blast wave can be evaporated
to gradually increase the density of the interior gas;
consequently, the SNR appears internally X-ray brightened (White \& Long 1991).
The fourth suggestion is that the mixed morphology is a projection effect.
For shell-like SNRs
that evolve in a density gradient such as at the edge of a molecular
cloud, if the line of sight is essentially aligned with the density
gradient as well as the magnetic field, the SNRs will appear as
thermal composites (Petruk 2001).
Here we report on high resolution \Chandra\ observation of 3C~391
and compare the properties with these scenarios.

\centerline{\includegraphics*[scale=0.4]{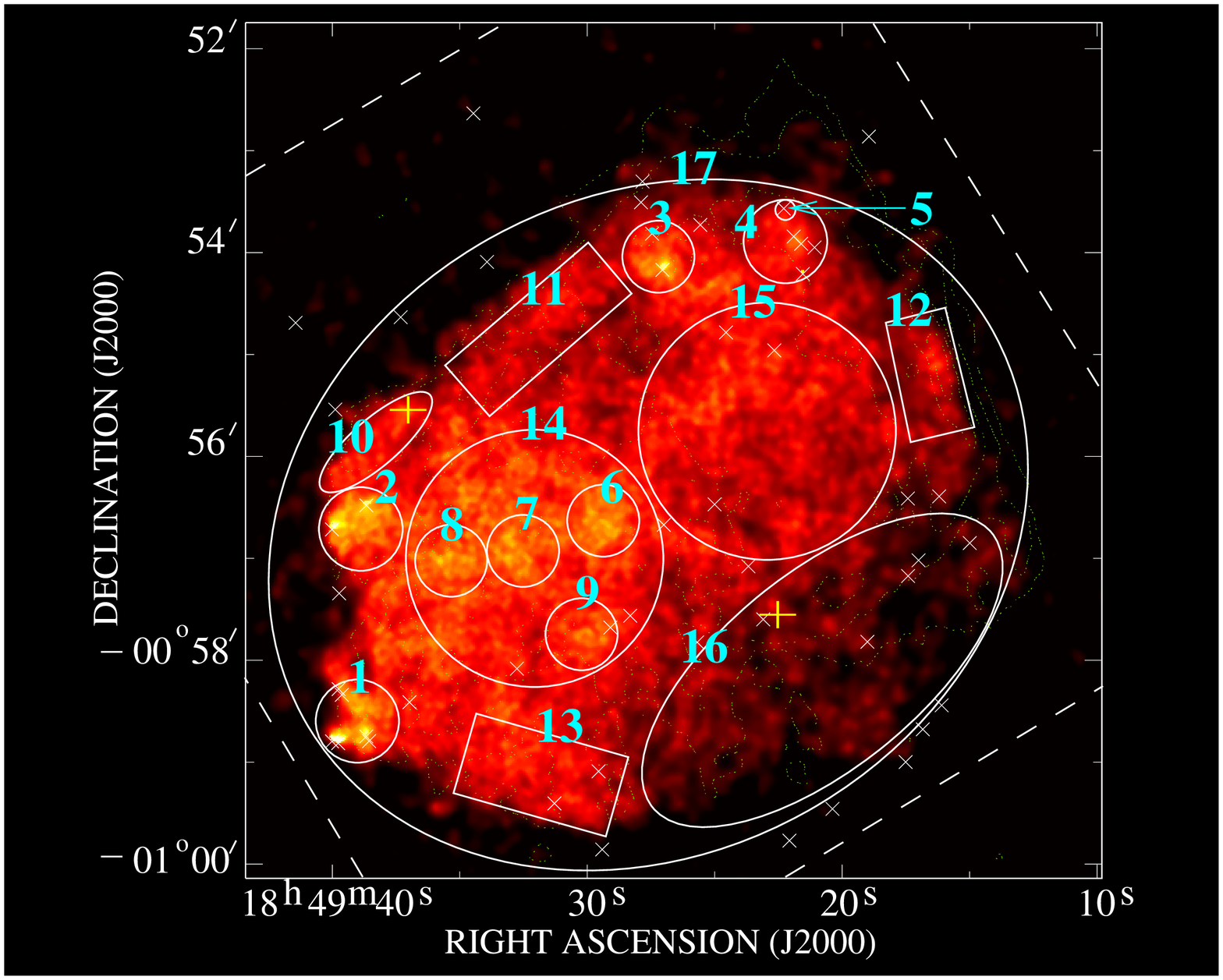}}
{\scriptsize
{\bf Fig.~1} 
Smoothed diffuse emission from SNR~3C~391 in the broad band
0.3-7.0 keV (with count-to-noise ratio of 6).
The color is logarithmically scaled in the range (1.19--331.85)$\E{-2}
{\rm~ct~s^{-1}~arcmin^{-2}}$.
The location of the sources removed from the data before
the smoothing are marked by white cross labels.
All the regions used for spectral analysis are indicated in white,
with cyan numerical labels.
The overlaid 1.5 GHz radio contours (in green) are at 1.5, 4.5, 13.6, 28.8,
and 50$\E{-3} {\rm~Jy~beam^{-1}}$ (Moffett \& Reynolds 1994).
The two yellow plus signs denote the OH maser points (Frail et al.\ 1996). 
The white dashed lines denote the border of the S3 chip.
}

\section{Data and analysis}\label{sec:img}
\centerline{\includegraphics*[width=135mm]{nar_band_sqrt.ps}}{\scriptsize
{\bf Fig.~2} Smoothed narrow band 1.2-1.5, 1.7-2.0, and 2.3-2.6$\keV$
(including Mg He$\alpha$, Si He$\alpha$, and S He$\alpha$, respectively)
diffuse emission images (with S/N ratio of 3)
overlaid with the dashed contours of 1.5 GHz radio emission
(at 1.5, 3.2, 8.25, 16.7, 43.7, and 62.2$\E{-3} {\rm~Jy~beam^{-1}}$)
(Moffett \& Reynolds 1994).
The seven levels of solid contours are plotted with square-root
intensity scales between the maximum
and the 15\% maximum brightness.
The two plus signs in each panel denote the OH maser points.
}
\vspace{5mm}

\centerline{\includegraphics*[width=135mm]{EW.ps}}{\scriptsize
{\bf Fig.~3} EW images of Mg, Si, and S lines. The Mg and Si (S)
images are extracted with $4"$ ($8"$) pixels and smoothed by a Gaussian
with $\sigma=12"$ ($16"$).
The seven levels of solid contours are plotted with square-root
intensity scales between the maximum and the 3\% maximum brightness.
The two plus signs in each panel denote the OH maser points.
}

SNR 3C~391 was observed with the \Chandra\ ACIS
on 03 August 2002.
The level 1 raw event data were reprocessed to generate a level 2
event file using the CIAO software package (version 2.3),
resulting a net 60.7 ks exposure.
In Figures 1 and 2, we show the diffuse X-ray map in 0.3-7 keV
and the narrow band 1.2-1.5,
1.7-2.0, and 2.3-2.6$\keV$ (including Mg He$\alpha$, Si He$\alpha$,
and S He$\alpha$, respectively) diffuse emission images.
The equivalent width maps (i.e., the line-to-continuum
ratio maps) for Mg, Si, and S are shown in Figure 3.

These X-ray images display the SE-NW elongated morphology,
revealing a highly clumpy structure of the remnant,
with clumps or knots located in both the SE and NW parts.
Several remarkable, very bright knotty features appear
on the east and SE border (regions \#1 and \#2) of 3C~391.
These bright knots on the border may be small clouds that have
recently been shocked by the supernova blast wave.
A bright enhancement is peaked near the NW border (region \#3).
A complex mixture of knots is seen in the SE part of the remnant interior,
including at least four bright enhancements indicated as
regions \#6, \#7, \#8, and \#9.

In Fig.1, arc- or shell-like structures are seen along the northeastern
and northern rim (regions \#10 and \#11).
An X-ray brightened slab at the west rim appears to be
very close to (just slightly behind) the radio peak emission (region \#12),
and may be related to a small dense region there.
On the southwest, faint diffuse emission seems to extend out of the
radio border.
On the corresponding location in the narrow-band and EW images
of Si and S lines (Figs.2b, 2c, 3b, and 3c),
there is a finger-like feature protruding radially out of the radio border.
This protrusion looks somewhat similar to the apparent protrusion
in Si and S lines that was recently revealed on the northeastern border
of Cas~A and explained as one of the jets of ejecta
(Hwang, Holt, \& Petre 2000; Hwang et al.\ 2004).
However, the small number of counts collected for this feature
makes it difficult to determine the metal abundances.

\noindent\begin{minipage}{80mm}
\centerline{\epsfig{figure=glo.ps,height=72mm,angle=270}}{\scriptsize 
{\bf Fig.~4} \Chandra\ ACIS spectra of the entire remnant of 3C~391 fitted with the VNEI model.
}
\end{minipage}\hspace{5mm}
\begin{minipage}{53mm}
With the point-like sources removed, X-ray spectra were extracted from
17 regions shown in Fig.1. 
Most of the small-scale regions are chosen to include the small features
of X-ray enhancement such as the knots and the faint shell like structures. 
The area on the S3 chip outside region \#17 was used for background.
\end{minipage}\\
The distinct line features, Mg~He$\alpha$ ($\sim1.35\keV$),
Si~He$\alpha$ ($\sim1.85\keV$), and S~He$\alpha$ ($\sim2.46\keV$)
in the spectra (see Fig.4) indicate the thermal origin of the diffuse gas.
We find that the spectra of the diffuse gas can be best described by the VNEI
model with the correction of interstellar absorption.
The spectral fit results are tabulated in the Table.

The absorption column density is found to generally increase across
the remnant from SE to NW,
consistent with the existence of
a molecular cloud to the NW (Wilner et al.\ 1998).
The spectral fits show that the diffuse emission from various regions
have ionization parameters ($n_e t$) close to or higher than
$10^{12}\cm^{-3}\s$.
This implies that the hot plasma in the SNR is very close to, or is
basically in, the ionization equilibrium.
The spectral fits also show that the diffuse emission from various
regions can be well fitted with solar abundances or
abundances very close to solar values.
The temperature of the gas interior to the SNR is generally 
$\sim$0.5-$0.6\keV$, with only small fluctuations.
The gas density of each defined region
was roughly estimated (as also listed in the Table).
Apart from the compact knots on the SE and eastern boundary,
most of the bright knots
have a gas density $\sim5$--$7f^{-1/2}\du^{-1/2}\cm^{-3}$
and most of the regions along the remnant border
have a density $\sim1$--$3f^{-1/2}\du^{-1/2}\cm^{-3}$
(where $\du=d/8\kpc$ denotes the distance).
The X-ray luminosity in 0.5--10 keV of the remnant
is $\sim3.5\E{36}\du^{2}\ergs\ps$.

{\scriptsize
\begin{tabular}{c|ccccc|c}
\multicolumn{7}{c}{VNEI fitting results with the 90\% confidence ranges
  and density estimates} \\ \hline\hline
regions &
$\chi^{2}/{\rm d.o.f.}$ & $\NH$ &
$kT_{x}$ & $n_e t$ &
$f\nel\nH V/\du^{2}$\hspace{2mm}$^{\rm a}$ &
$\nH/f^{-1/2}\du^{-1/2}$ \\
    &   &
($10^{22}\cm^{-2}$) &
(keV) & ($10^{11}\cm^{-3}\,{\rm s}$) &
($10^{57}\cm^{-3}$) 
& (cm$^{-3}$)
 \\ \hline
1  & 90.5/65 & $2.8\pm0.1$ &
  $0.67^{+0.02}_{-0.04}$ & $>200$ & $2.55^{+0.27}_{-0.53}$ & 18\\
2  & 87.0/79 & $2.7\pm0.1$ &
  $0.58^{+0.06}_{-0.05}$ & $5.1^{+4.6}_{-1.3}$ &
  $3.84^{+1.33}_{-0.89}$ & 11\\
3  & 68.8/48 & $3.4\pm0.2$ &
  $0.56^{+0.06}_{-0.09}$ & $>6.4$ & $3.46^{+2.08}_{-0.90}$ & 6.4\\
4$^{\rm b}$  & 108.9/79 & $4.1^{+0.3}_{-0.2}$ &
  $0.62^{+0.04}_{-0.05}$ & $>30$ & $9.42^{+3.10}_{-2.33}$ & 4.8\\
      &    & \multicolumn{4}{c}{(
  {\bf [Mg/H]=$1.23^{+0.69}_{-0.44}$, [Si/H]=$0.66^{+0.19}_{-0.15}$,
   [S/H]=$0.51^{+0.22}_{-0.20}$})} \vline  \\
6  & 65.2/58 & $2.7^{+0.2}_{-0.1}$ &
  $0.56^{+0.08}_{-0.06}$ & $5.0^{+3.8}_{-2.1}$ &
  $3.20^{+1.41}_{-0.98}$ & 6.1\\
7  & 57.8/58 & $3.0^{+0.1}_{-0.2}$ &
  $0.54^{+0.05}_{-0.06}$ & $>5.5$ & $3.97^{+1.65}_{-0.98}$ & 6.8\\
8  & 72.4/52 & $2.9^{+0.2}_{-0.1}$ &
  $0.63^{+0.08}_{-0.07}$ & $>3.4$ & $2.52^{+0.60}_{-0.65}$ & 5.4\\
9  & 58.2/43 & $2.9\pm0.2$ & $0.63^{+0.07}_{-0.11}$ &
  $>2.7$ & $1.72^{+1.35}_{-0.41}$ & 4.5\\
10  & 44.2/44 & $3.0^{+0.2}_{-0.1}$ &
  $0.79^{+0.14}_{-0.10}$ & $3.1^{+3.6}_{-1.4}$ &
  $1.16^{+0.52}_{-0.34}$ & 2.3\\
11  & 96.9/72 & $2.8\pm0.1$ &
  $0.59^{+0.06}_{-0.04}$ & $>3.8$ & $3.17^{+0.38}_{-0.77}$ & 1.9\\
12  & 49.8/32 & $3.7^{+0.6}_{-0.4}$ &
  $0.58^{+0.11}_{-0.12}$ & $>265$ & $2.28^{+3.19}_{-0.96}$ & 2.0\\
13  & 107.0/82 & $3.0^{+0.2}_{-0.1}$ &
  $0.46^{+0.04}_{-0.03}$ & $>106$ & $8.23^{+2.73}_{-2.12}$ & 2.6\\
14$^{\rm b}$  & 273.6/180 & $2.9\pm0.1$ &
  $0.55\pm0.02$ & $>9.0$ & $35.00^{+3.75}_{-5.38}$ & 2.9\\
      &    & \multicolumn{4}{c}{(
  [Mg/H]=$1.12^{+0.13}_{-0.09}$, [Si/H]=$0.87^{+0.08}_{-0.05}$,
   [S/H]=$0.80^{+0.13}_{-0.09}$)} \vline \\
15$^{\rm b}$  & 237.8/163 & $3.5\pm0.1$ &
  $0.54^{+0.02}_{-0.01}$ & $>23$ & $34.08^{+5.42}_{-4.20}$ & 2.8\\
      &    & \multicolumn{4}{c}{(
  [Mg/H]=$0.92^{+0.14}_{-0.13}$, [Si/H]=$0.53\pm0.05$,
   [S/H]=$0.64^{+0.12}_{-0.11}$)} \vline \\
16$^{\rm b}$  & 156.3/127 & $3.2\pm0.2$ &
  $0.53^{+0.04}_{-0.06}$ & $>239$ & $10.0^{+5.42}_{-2.60}$ & 1.1\\
      &    & \multicolumn{4}{c}{(
  [Mg/H]=$0.93^{+0.28}_{-0.26}$, [Si/H]=$0.54^{+0.12}_{-0.11}$,
   [S/H]=$0.54^{+0.24}_{-0.21}$)} \vline \\
17$^{\rm b}$  &  710.5/296 & $3.1\pm0.1$ &
  $0.56\pm0.01$ & $>12.8$ & $150\pm10$ & 1.9\\
      &    & \multicolumn{4}{c}{(
  [Mg/H]=$0.97^{+0.07}_{-0.05}$, [Si/H]=$0.70\pm0.03$,
   [S/H]=$0.71^{+0.06}_{-0.05}$)} \vline & \mbox{} \\ \hline
\multicolumn{7}{l}{a: $f$ is the filling factor of the hot gas.}\\
\multicolumn{7}{l}{b: Making abundances of Mg, Si, {\bf and S} free
parameters apparently improves the fit.}
\end{tabular}
}

\section{Discussion on the composite appearance}

Here we briefly compare the four mechanisms mentioned in \S1 with
the properties found from our spatially-resolved spectral analysis.

i) {\em Projection Effect} The molecular cloud is located in the NW but
the X-ray emission is enhanced not only in the NW half, but also
in the SE half. Additionally,
the variation of the hydrogen column density across the remnant
implies that the density gradient of the ambient medium seems to
be close to the projection plane. Therefore,
the projection effect does not match the observational properties.
ii) {\em Radiative Rim} 
The filamentary near-infrared
[Fe II] and the mid-infrared 12-18 $\mu$m [Ne II] and [Ne III] emission
along the NW radio shell provide some evidence for a radiative cooling at
the rim (Reach, Rho, \& Jarrett 2002).
Along the NW border, the shell formation time is
$t_{\rm shell}\approx3.8\E{3}\Eu^{3/14}\yr$,
comparable to remnant's age $t=(2/5)(\rs/\vs)\sim4\E{3}\yr$.
However the X-ray emission along the border (e.g.\ regions \#10, \#11,
\#12, and \#16), which arises from hot gas ($\sim7\E{6}\K$), 
indicates that a considerable amount of gas at the blast shock has
not yet suffered significant radiative cooling.
%
Additionally, the rim cooling mechanism can not explain what
the central bright clumpy emission is.
iii) {\em Thermal Conduction}
For 3C~391, the thermal conduction scenario 
is favored by the radiative filaments along the NW border and
the high ionization timescale implicative of relatively little
newly shocked material.
The conduction timescale
$t_{\rm cond}\sim 5.2\E{3}(\ell/2.3\parsec)^2\yr$
for the spatial scale $\ell$ of order the separation between clumps
(typified by $1'\sim2.3\parsec$) is comparable to the remnant's age
and hence implies a role of conduction in smoothing the
interior temperature profile.
However, the thermal conduction scenario asks for a decrease of temperature
and an increase of gas density with remnant radius.
This predicted behavior is inconsistent with the uniform distribution
of the gas temperature and density.
iv) {\em Cloudlet Evaporation}
The relatively uniform distribution of temperature (even with slightly
lower values at the center) is actually expected by the
White \& Long (1991) cloudlet evaporation model for model parameters
$\tau\rightarrow\infty$ and $C/\tau\gsim3$,
where $\tau$ is the ratio of the cloud evaporation timescale to the
SNR's age and $C$ is the ratio of the mass in the cloudlets
to the mass of ICM.
The ratio between the mean density ($\sim2f^{-1/2}\du^{-1/2}\cm^{-3}$)
and the density along the border ($\sim1$--$3f^{-1/2}\du^{-1/2}\cm^{-3}$)
 is basically
consistent with that predicted in the evaporation model.
In fact, the highly clumpy structure unveiled in this observation
lends support to the conjecture that the ambient molecular cloud is
inhomogeneous.
A combination of very dense clumps and moderately dense gas are directly
observed in the millimeter molecular lines (Reach \& Rho 1999).
The cloudlets engulfed by the supernova blast wave can act as a large
reservoir of interior gas by gradual evaporation. This could also
explain why the interior gas density
is much lower than that of the ambient cloud gas ($\sim30$-$100\cm^{-3}$).
With this model, the supernova explosion energy is
$E\sim0.3$--$1.4\E{51}\du^3(\no/0.3\cm^{-3})\ergs$.
The main difficulty with the model is the high ionization age
($>10^{12}\s$) in some regions (e.g.\ region \#1) opposed to the
low age of the newly evaporated gas.


\end{document}